\documentclass[longbibliography,
 reprint,
 amsmath,amssymb,
 aps,
 pre,
]{revtex4-2}

\usepackage{graphicx}
\usepackage{dcolumn}
\usepackage{bm}
\usepackage{braket}
\usepackage{color}

\begin{document}

\preprint{APS/123-QED}

\title{Preparation and observation of jammed particles with power size distribution}

\author{Daisuke S. Shimamoto}
\email{shimamoto-daisuke806@g.ecc.u-tokyo.ac.jp}
\author{Miho Yanagisawa}%
 \email{myanagisawa@g.ecc.u-tokyo.ac.jp}
 \altaffiliation[Also at ]{Center for Complex Systems Biology, Universal Biology Institute, The University of Tokyo., Graduate School of Science, The University of Tokyo, Hongo 7-3-1, Bunkyo, Tokyo 113-0033, Japan}
\affiliation{%
Komaba Institute for Science, Graduate School of Arts and Sciences, The University of Tokyo, Komaba 3-8-1, Meguro, Tokyo 153-8902, Japan
}%

\date{\today}

\begin{abstract}
Several materials, such as rocks, powders and molecules, are multi-component systems. However, compared to single-component systems, it is difficult to understand the physical component. In this study, as a coarse-grained model for powders with extremely large size variations, we experimentally and numerically filled circular particles with a power size distribution and investigated their structure at the jamming transition point. In the experiment, oil in water droplets following a power size distribution were created, and then we constructed a model with steady injection and fracture to explain the size distribution. In numerical calculations, the dependence of the packing structure on the exponent of the particle size distribution was investigated. The existence of fractal structure with cutoffs was experimentally and numerically found from the structure factor. Numerical calculations show that the area fraction of rattlers is almost independent of exponent, but the number fraction is significantly dependent.
\end{abstract}

\maketitle

\section{introduction}

In the realm of theoretical statistical physics, research has predominantly concentrated on the development and analysis of simplified models. These models, characterized by uniform elements with identical interactions or a limited variety of components, have been instrumental in elucidating the underlying principles of complex phenomena. For example, the Ising model, which only places spin variables on the lattice, qualitatively reproduces the phenomena exhibited by magnetic materials, and the Flory-Huggins theory, which places chains and balls on the lattice, predicts the physical quantities of mixtures of polymers and solvents. Thus, many theories have quite succeeded in predicting physical quantities of matter composed of a few types of elements.

The scope of statistical physics is expanding beyond these elementary models to encompass systems of greater complexity. There are various natural components, distinguishable by their interactions, responses to external field, mass, and more. Even when focusing solely on pairwise interactions, the systems have complexity owing to variable dependencies on distance and angle. This complexity poses challenges in formulating effective coarse-grained models that accurately represent real-world particles. Consequently, researchers often resort to studying the simplest conceivable systems, such as monodisperse particles with isotropic interactions, or their slightly more complex counterparts like particles with small size polydispersity\cite{yanagisawa2021size,desmond2013experimental} or the Kob-Anderson mixture, which is a mixture of particles with a radius ratio of 1:1.4\cite{o2003jamming,kob1995testing,o2002random,zhang2009thermal,iikawa2016sensitivity}.

Increasing the particle size ratio\cite{biazzo2009theory,berthier2009glass,ikeda2021multiple,hara2021phase} leads to the presence of two distinct phases and phase transition-like behavior between them. In addition, the introduction of surface roughness allows the consideration of "monodisperse particles with anisotropic interactions". Experiments in quasi-2D systems using slightly non-spherical colloids have reported that spatial anisotropy of interactions alters mechanical properties\cite{lootens2005dilatant,hsu2018roughness} and induces self-assembly\cite{sacanna2011shape} and gelation owing to capillary forces\cite{kato2023surface}. The vibrational properties of 2D particles with slightly rough surfaces are analyzed using numerical calculations and mean-field theory\cite{ikeda2019mean}. These studies show that the extension of theoretical models for single-component systems can be used to explain an increasing number of phenomena for multicomponent systems.

The extension of the theoretical model from single-component systems has increased the number of theoretically explainable phenomena in multicomponent systems, while most granular materials we encounter are much more polydisperse and multicomponent, than two-dispersion system mentioned above. Thus, there is a significant gap between idealized, simple systems and the complex systems around us, highlighting the considerable challenge of dealing with polydisperse, multicomponent systems.

Addressing the complexity of granular systems necessitates considering broad size distributions, research has been performed to understand systems composed of particles whose radii follow a power-law distribution. These systems, prevalent across various natural phenomena from high-energy impact or shear fracturing to steady processes of injection\cite{oddershede1993self,ishii1992fragmentation,katsuragi2004crossover,kobayashi2021fragmentation, ben2010force} and aggregation\cite{hayakawa1987irreversible,takayasu1988power}, offer insightful models for the actual systems. For instance, power-law distributions serve as models for explaining the size distribution of regolith on planetary surfaces\cite{grott2020macroporosity} or fault gauge\cite{sammis1987kinematics}, as well as for understanding the behavior of metal particles\cite{singh2012study} and nanoparticles within cells\cite{alexandrov2022dynamics}.

In the context of exploring granular systems characterized by broad size distributions, significant attention has been directed towards particles whose radii follow a power-law distribution. Such investigations, both theoretical and numerical, have unveiled that these systems exhibit fractal characteristics as well as are capable of fill space without voids, when $a$ falls within a specific range, $d_{\rm AP}+1\leq a<d+1$ where $d$ is the spatial dimension and $d_{\rm AP}$ is the fractal dimension of the surface of an Apollonian packing in $d$-dimensional space. Note that the influence of cutoffs at the maximum and minimum radii present in the actual particle size distributions. Even in coarse-grained models, maximum and minimum values in the size distribution are necessary to avoid divergence in particle number and volume. A distribution that remains continuous between two cutoff lengths and is zero elsewhere is determined by two parameters: the exponent $a$ and the ratio of the cutoffs. We have reported the structure at the jamming transition point of two-dimensional circular particles following power-law size distributions from both experimental\cite{shimamoto2023common} and numerical perspectives. For example, we observed that the contact network maintains a consistent structure across variations in $a$, provided $a<3$. This property is attributed to the fractal nature of the distributions\cite{shimamoto2023common}. Moreover, within the interval $2<a<3$, our research has indicated a notably higher packing fraction at the jamming transition point relative to systems comprised of monodisperse or bimodal distributions. Recently, Kim and Ikeda \cite{kim2024replica} performed replica calculations on glass with infinite dimensions following exponential size distributions (which cleverly correspond to power-law distributions in finite dimensions by depending on dimension $d$ and revealed that near $a=d$, it does not exhibit glass transitions even at high densities. Additionally, they unveiled the existence of a partial glass phase wherein only larger particles transition into a glassy state, while smaller particles retain their liquid state for $a<d$. Additionally, numerical investigations have been conducted on the structure factor of systems filled with particles following power-law distributions using various methods in systems ranging from one to three dimensions\cite{cherny2023dense}. However, it remains unclear how the structure at at the jamming transition point depends on the power-law exponent $a$.

To understand the behavior of granular systems with power-law size distributions from the perspective of statistical mechanics, experimental research is essential alongside numerical simulations and theoretical analysis. The challenge of controlling the polydispersity of spherical or circular particles remains substantial. Although microfluidic devices have facilitated the generation of droplets and bubbles with specific size distributions\cite{zhu2017passive}, these polydisperse materials have yet to be extensively applied in the empirical validation of statistical models. Our previous experiments with oil droplets following power-law size distributions did not fully elucidate the underlying formation mechanisms. Consequently, this study aims to clarify the processes generating of particles with power-law size distributions, using a model with steady destruction and injection. Additionally, we investigate how the two-dimensional packing structure of these oil droplets evolves with variations in the power-law exponent $a$, thereby enhancing our understanding of granular systems from both a theoretical and practical perspective.

\section{Materials and methods}
\subsection{Materials}
UltraPure DNase/RNase-free distilled water (Invitrogen, Waltham, MA) containing Tween 20 surfactant (Sigma-Aldrich, St. Louis, MO) was used as the continuous aqueous phase. Mineral oil (Nacalai Tesque, Kyoto, Japan) containing Span 80 surfactant (Tokyo Chemical Industry, Tokyo, Japan) was used as the dispersed oil phase. A lipophilic dye, capsanthin in vegetable oil (Tokyo Chemical Industry), was also added to the mineral oil to improve the visibility of the oil-in-water (O/W) droplets.

\subsection{Preparation of droplets}

We used O/W droplets stabilized by a surfactant as a model of soft and frictionless particles \cite{desmond2013experimental}. The O/W droplets were prepared using an aqueous solution containing 1 wt\% Tween 20 and a mineral oil containing 0.1 wt\% Span 80 and 2 wt\% capsanthin oil. See our previous paper for details on materials and experimental methods \cite{shimamoto2023common}. 

\begin{figure}
    \centering
    \includegraphics[width=86mm, bb = 0 0 921.827121 174.110779]{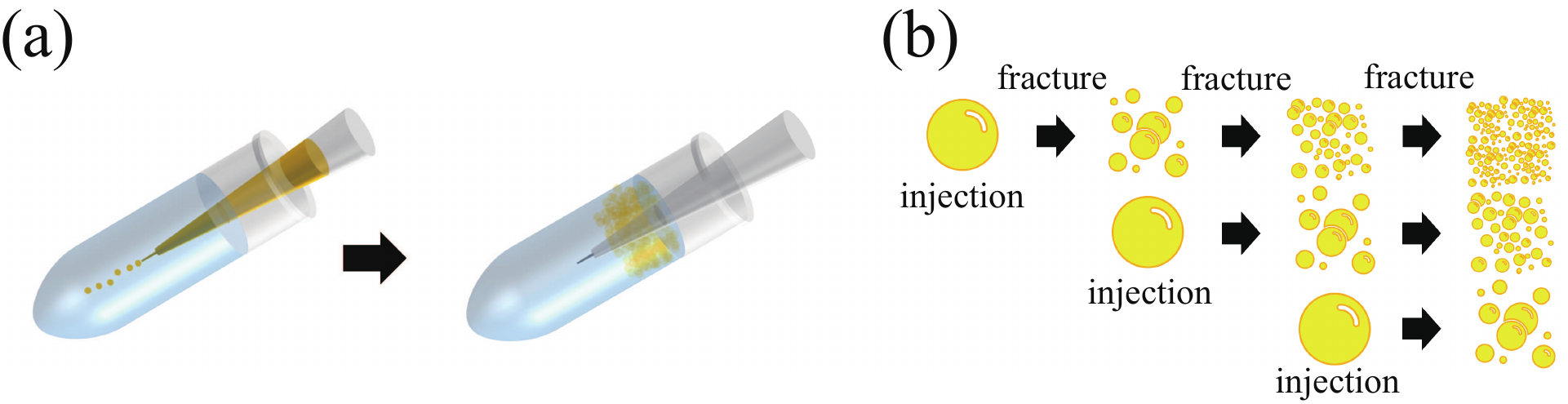}
    \caption{(a) Schematic of the microfluidic device used to produce monodisperse and bidisperse oil droplets (yellow) in water (blue). (b) Schematic of the procedure used to produce polydisperse oil droplets (yellow) by repeatedly injecting the oil phase into the water and fracturing the oil droplets.}
    \label{fig3}
\end{figure}

Shortly, the bidisperse O/W droplets were prepared by a centrifugal microfluidic device (Fig.\ref{fig3} (a)). The device consists of three components: a glass capillary, micropipette tip, and microtube. The glass capillaries with an inner diameter of less than $\sim 50\,\mu$m at the thin tip and a length of 8 mm were fabricated from a capillary (G-1; Narishige, Tokyo, Japan) using a puller (PC-10; Narishige) and a microforge (MF-900; Narishige). These capillaries were attached to the end of micropipette tips (200 $\mu$L standard tip; Labcon, Petaluma, CA). The micropipette tip was filled with $80 \,\mu$L of the mineral oil and attached to the microtube containing $500\,\mu$L of the aqueous solution by passing through a 6-mm-diameter hole carefully drilled in the lid of the microtube. The device was centrifuged for 1 minute using a tabletop centrifuge.
 The size distribution of the O/W droplets depends on the size and shape of the glass capillary. A capillary with a smooth tip of 25 $\mu$m inner diameter produces monodisperse droplets and a capillary with a chipped tip of 50 $\mu$m inner diameter produces bidisperse droplets.
 
 Polydisperse O/W droplets are prepared by repeatedly (i) injecting the oil phase into microtubes containing the aqueous solution and (ii) fracturing the oil phase by impact (Fig.\ref{fig3} (b)). A small amount (approximately 30 $\mu$L) of the oil phase is injected into a microtube containing approximately 800 $\mu$L of the aqueous solution. The impact of tapping and the resulting complex flow break up the oil phase to form small oil droplets. This process is repeated X times to produce polydisperse droplets.
 
 The resulting W/O droplets are deformed into a pancake shape by sandwiching them between two glass plates separated by double-sided tape (thickness $\sim$0.09 mm). The radius of the prepared droplets ranges from $14$ to $421$ $\mu$m.

\subsection{Observation and image analysis}
The images of the two-dimensional droplets were captured using a camera (a2A5328-15ucPRO; Basler) attached to a microscope (SZX16). Images were analyzed using ImageJ (\url{http://rsb.info.nih.gov/ij/}), free software from the National Institutes of Health (NIH). Droplets in the images were detected by binarization after noise removal using a median filter and a fast Fourier transform (FFT) bandpass filter. As shown in the microscopic image in Fig. \ref{disp_exp}(a), the water outside the droplets appears blue, while the surface of the oil droplet with a large refractive index difference appears as a black ring.

\subsection{Numerical calculations}

Numerical calculations were performed using molecular dynamics simulations. The packing of particles at the jamming transition point for the two-dimensional system was prepared in the same way as in the previous paper \cite{shimamoto2023common}. See also \textcolor{black}{\cite{hara2021phase}} for the procedures. In short, the particle radius $r$ was randomly set according to a power distribution. 
The particles' radii followed power distribution with the exponent $a$ smaller than 3, and $R=r_{\rm max}/r_{\rm min}=50$.
Note that fixing the number of particles $N$ at 4000 reduces the size range produced when $a$ is large. Therefore, we set the number of particles to be sufficiently large, i.e., 4000 for $a<2$, 8000 for $2\leq a\leq 2.8$, and 16000 for $a=3$. 
\textcolor{black}{The interaction between particles follows a purely repulsive harmonic potential. This is a pairwise potential between elastic two-dimensional disks \cite{landau1986theory}.}

\section{Results and Discussion}

\subsection{Experimental compression of droplets}
\begin{figure}
    \centering
    \includegraphics[width=8.6cm, , bb = 0 0  651 424]{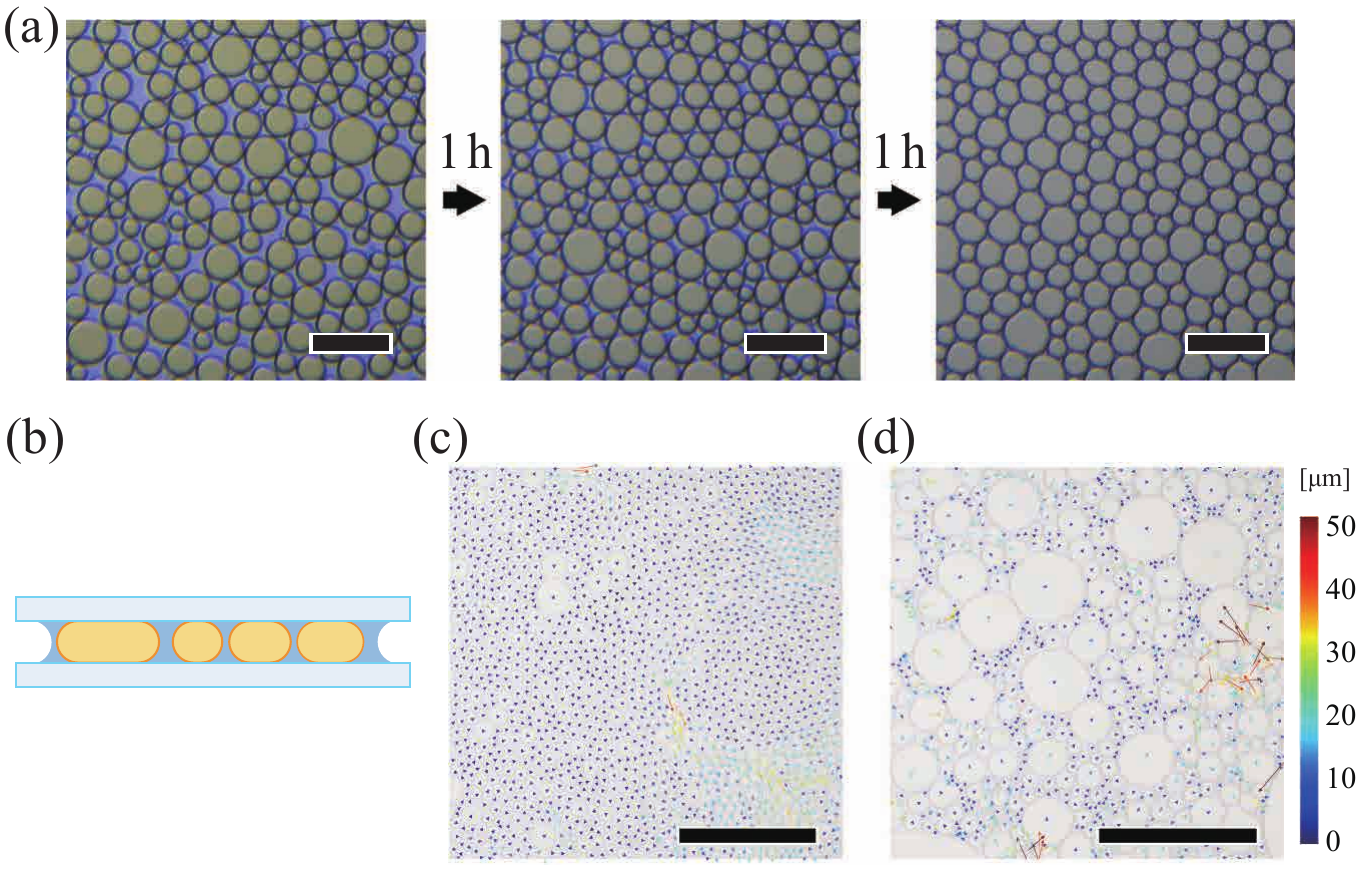}
    \caption{(a) Increase in the packing fraction of the oil droplets due to evaporation of the outer aqueous phase. One hour elapsed between the acquisition of the two adjacent images. The scale bars are 200 $\mu$m.
    (b) Schematic illustration of the vertical cross section of the experimental setup. The yellow part and the blue part are oil and aqueous phase, respectively. The light blue rectangles are glass plates.
    Displacement of (c) bidisperse droplets and (d) polydisperse droplets during 10 s. This displacement was magnified 10 times and overlaid on the microscopic image as arrows. The color of the arrows indicates the magnitude of the displacement. The scale bars are 1 mm.
    }
    \label{disp_exp}
\end{figure}

To analyze the structure of highly packed droplets near the jamming transition point, $\phi_{\rm J}$, the packing fraction of the oil droplets was increased by the evaporation of the aqueous phase outside the droplets. As shown in the schematic diagram in Fig. \ref{disp_exp}(b), the edge of the aqueous phase containing oil droplets is in contact with air. Because evaporation occurs in the aqueous phase but not in the oil phase, the packing fraction occupied by O/W droplets in the region bounded by the air-water interface increases with time. The average rate of the packing fraction increase was 5$\%$/h. The packing fraction increased from below to above jamming transition point $\phi_{\rm J}\simeq 0.84$ \cite{o2002random}. No coalescence or splitting of droplets was observed throughout the process.

When the packing fraction exceeds $\phi_{\rm J}$, the circular droplets contact each other and deform, forming a flat contact line (Fig. \ref{disp_exp}(a)). The dynamics of the droplets as the packing fraction increases seem to depend on the droplet size distribution. In the bidisperse system, the displacement of the droplets propagated over a long distance as a band (Fig. \ref{disp_exp}(b)). Such deformations have been also reported in simulations \cite{maloney2006amorphous} and experiments \cite{yanagisawa2021size}.
However, in the polydisperse system, band displacement as well as a rotational motion of multiple droplets were observed (right side of Fig. \ref{disp_exp}(c)). This dynamics of polydisperse systems with increasing packing fraction is interesting; however, it is outside the scope of the present paper and will be left for future research.

\subsection{Droplet size distribution and the manipulation}

\begin{figure}
    \centering
    \includegraphics[width=8.6cm, , bb = 0 0 256.657440 155.521835]{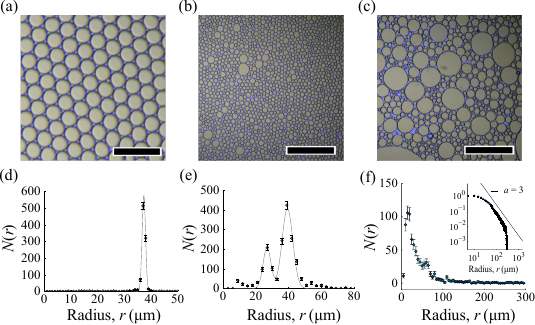}
    \caption{(a-c) Microscopic images of O/W droplets packed in a quasi-two-dimensional cell. The droplet size is monodisperse (a), bidisperse (b), and polydisperse with a power size distribution, respectively. The scale bars are 200 $\mu$m for (a), and 1 mm for (b,c). (d-f) Droplet radius distribution, $N(r)$. (d) In the monodisperse system, $N(r)$ is concentrated around the mean of 37 $\mu$m  (standard deviation $\sim$4). (e) In the bidisperse system, the peaks are at the mean of 29 $\mu$m and 39 $\mu$m. (f) $N(r)$ follows a power size distribution of a $\sim$3 in the region of one order of magnitude. Figs. (b,c,e,f) are reprinted from \textcolor{black}{our previous paper \cite{shimamoto2023common}}.}
    \label{crystallize}
\end{figure}

The simplified centrifugal microfluidic device produced monodisperse or bidisperse droplets (Fig. \ref{crystallize}(a, b, d, e)). Bidisperse droplets, which have two distinct peaks in the histogram of the droplet size distribution, were prepared for comparison with polydisperse systems. The reason for using bidisperse droplets instead of monodisperse droplets is to avoid crystallization, which is characteristic of two-dimensional systems (Fig. \ref{crystallize}(a, d)). 
The diameter and the smoothness of the tip of the capillary, built in the simplified microfluidic device, determine the size distribution of the droplets \cite{zhu2017passive}. A capillary with a smooth tip of 25 $\mu$m inner diameter produced monodisperse droplets \textcolor{black}{(Fig. \ref{crystallize}(a, d))}, and a capillary with a chipped tip of 50 $\mu$m inner diameter produces bidisperse droplets \textcolor{black}{(Fig. \ref{crystallize}(b, e))}. A large tip diameter increases the mean and variance of the droplet size, and a chipped tip produces bidisperse droplets, contrary to monodisperse droplets that result from a smooth tip. 

In the case of polydisperse droplets, the procedure of adding a small amount of oil to a tube half-filled with water, and then applying an impact was repeated. The size distribution of the obtained polydisperse droplets ranges over $10$ times the power distribution with an exponent of $-3$ (Fig.\ref{crystallize}(c, f)). The principle that led to the $a=3$ droplet size distribution is discussed here based on the repeated sequential fracture and injection.
It is expected that added droplets will be repeatedly broken up by the complex flow in the tube with each impact of the tap. We assume that this breaking (or fracturing) is a random multiplicative process. Because repeated fractures are known to produce lognormally distributed debris (theory:\cite{kolmogorov1941logarithmically}, experiment:\cite{kobayashi2006statistical}), the probability distribution $P_m(V)$ of the droplet volume $V$ is lognormal:
\begin{equation}
    P_m(V)=\frac{1}{\sqrt{2\pi m\sigma^2}V}\exp\left(-\frac{\left(\log V-m\mu\right)^2}{2m\sigma^2}\right),
\end{equation}
where $\mu$ and $\sigma$ are constants that characterize each fracture process and $m$ is the number of fractures experienced by the droplets after each impact (See Appendix \ref{ap_lognormal} for the derivation of this distribution.).
From the above, the size distribution of the oil droplets resulting from the fracture of the oil added in an injection is obtained. Subsequently, we sum the size distribution of the oil droplets produced by the oil phase added in multiple injections. Therefore, it is necessary to obtain a frequency distribution of the absolute number of droplets instead of a normalized probability distribution, thus we convert a probability distribution, $P_m(V)$ to a frequency distribution, $N_m(V)$:
\begin{eqnarray}
    N_m(V)&=&\frac{P_m(V)}{\int^\infty_0 dV VP_m(V)},\\
    &=&\frac{1}{\sqrt{2\pi m\sigma^2}V}\nonumber\\
    &&\exp\left(-\frac{\left(\log V-m\mu\right)^2}{2m\sigma^2}-m\mu-\frac{m\sigma^2}{2}\right). \label{lognormalNV}
\end{eqnarray}
This is because the total volume of particles before and after the fracture process must be conserved, and the frequency distribution of $V$ is the the probability distribution of $V$ divided by the mean of $V$. 

Considering the logarithm of both sides, it can be organized as follows:
\begin{eqnarray}
    \log N_m(V)&=&-\frac{\left(\log V\right)^2-2m\mu \log V+m^2\mu^2}{2m\sigma^2}\nonumber\\
    &&-m\mu-\frac{m\sigma^2}{2}-\log V\nonumber\\
    &&-\frac{1}{2}\log 2\pi m\sigma^2,\\
    &=&-\frac{1}{2m\sigma^2}\left(\log V-m\left(\mu-\sigma^2\right)\right)^2\nonumber\\
    &&-2m\mu-\frac{1}{2}\log 2\pi m\sigma^2.
\end{eqnarray}
The vertices of the parabola that appear when $N_m(V)$ is plotted bilogarithmically are ($m(\mu-\sigma^2)$, $-2m\mu-1/2\log 2\pi m\sigma^2$), which vertices lie on a line with slope $-2$. In addition, because the width of the parabola is proportional to $\sqrt{m}$, the valley between the peaks disappears as $m$ increases.
We now can obtain a volume distribution, $N(V)=\sum_m N_m(V)\propto V^{-2}$.
Subsequently, the volume distribution $N(V)\propto V^{-2}$ is converted into a distribution $N(r)$ of radius $r$:
\begin{eqnarray}
    N(r)&\propto& V^{-2}\frac{dV}{dr},\\
    &\propto&\left(r^d\right)^{-2}r^{d-1},\\
    &\propto&r^{-d-1}, \label{-d-1}
\end{eqnarray}
where the spatial dimension was set to $d$.
Because the droplets used in the experiments in this study are quasi-two-dimensional, so substituting $d=2$ into Eq.\ref{-d-1}, we obtain $N(r)\propto r^{-3}$, which is in agreement with the experimental result (Fig.\ref{crystallize}(f)).

\subsection{Structure factor}

We have previously reported on the contact network of packed droplets at the jamming
transition point\cite{shimamoto2023common}.
Here, we analyze it in more detail based on the structure factor, which is calculated using the following equation:
\begin{eqnarray}
    S({\bf k})=\frac{1}{N}\sum_{i,j} {\rm e}^{i{\bf k}\cdot({\bf x}_i-{\bf x}_j)},
\end{eqnarray}
where ${\bf k}$ and ${\bf x}_i$ denotes wave-number vector and position vector of $i$-the droplet, respectively. In the experiment, the structure factor was calculated from the positions of the center of the droplets detected from the experimentally obtained images.

\begin{figure}
    \centering
    \includegraphics[width=8.6cm, , bb = 0 0 395 321]{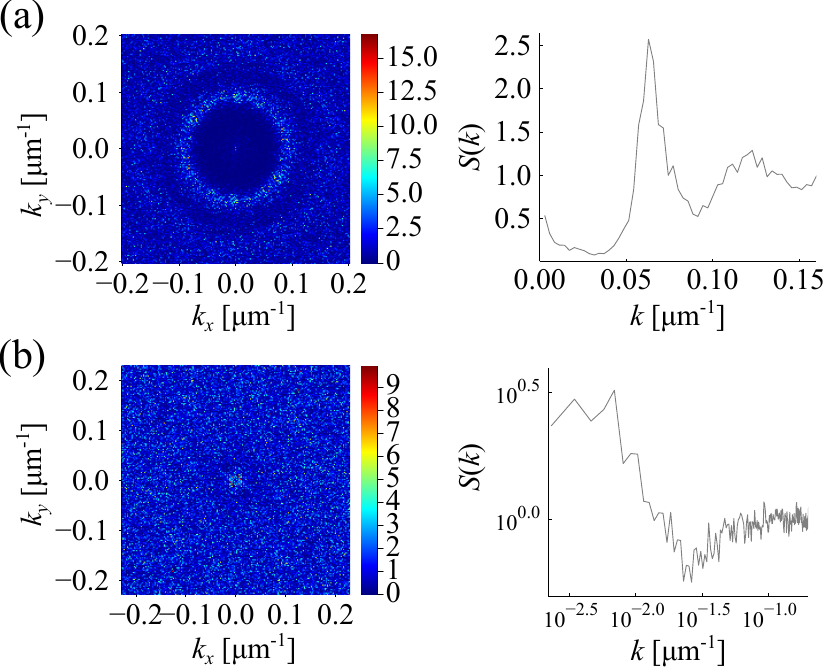}
    \caption{\textcolor{black}{Structure factor (left) and their projection to the radial direction, S(k)} (right) for (a) bidisperse and (b) polydisperse O/W droplets. The characteristic length scale ($\simeq 90$ $\mu$m) appears in the bidisperse system. The polydisperse system shows a flat pattern indicating the absence of the characteristic length scale.}
    \label{Sk_exp}
\end{figure}

The structure factor of the bidisperse system shows multiple concentric circles (Fig.\ref{Sk_exp} (a)), while the polydisperse system is uniform except for the bright region near the origin (Fig.\ref{Sk_exp})(b)).
\begin{figure}
    \centering
    \includegraphics[width=8.6cm, , bb = 0 0  488 337]{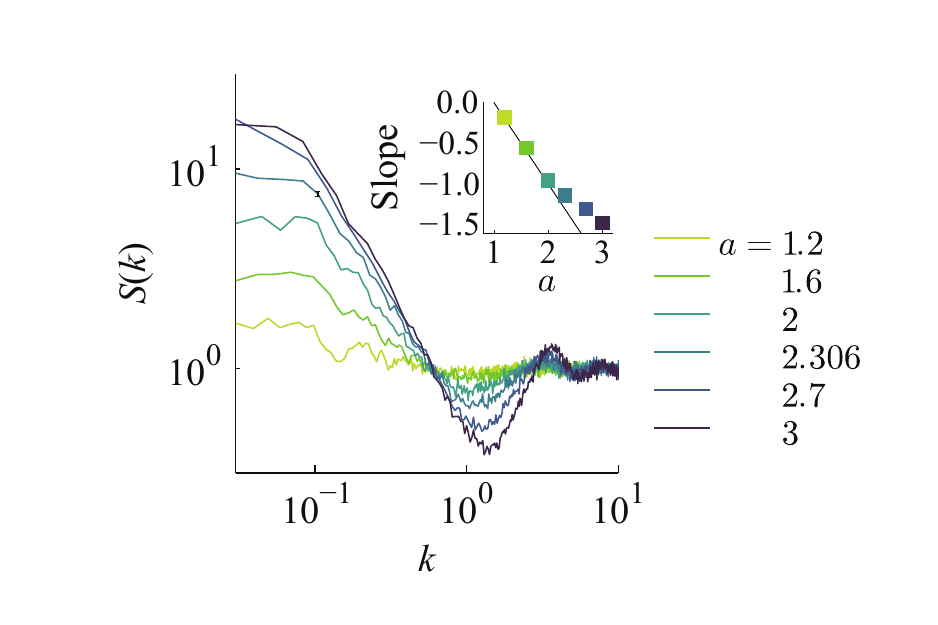}
    \caption{
    Numerically obtained structure factors for packed polydisperse particles with $a=1.2, 1.6, 2, 2.306, 2.7$, and $3$. Length scales were normalized by the radius of the smallest particle in each system; averages of 30 trials are shown. For visibility, only one representative standard error is shown as an error bar. \textcolor{black}{For all $a$, a transition was observed from the plateau region to the power region with increasing wavenumber $k$, followed by peaks.}
    The inset shows a plot of the slope of linear fits of double log plots \textcolor{black}{of the power region} of the structure factor versus $a$. The black line is a prediction based on the previous result about the packing of power-sized particles with dense and non-jamming way \cite{cherny2023dense}.
    }
    \label{Sk}
\end{figure}

The structure factor of the bidisperse system is similar to those reported for liquid or supercooled liquid \cite{kob1999computer}, i.e. they show spatial isotropy, in contrast to crystals, and the presence of several specific length scale structures at short distances.
The results for polydisperse systems show isotropy and the absence of crystals as in bidisperse systems as well as imply the absence of a distinct representative length scale.

To understand the difference between the structure factors of polydisperse and bidisperse systems, the structure factors were obtained in the same way using simulation results of polydisperse systems with exponents $a$ varying between 1.2 and 3. The numerically obtained structure factor is shown in Fig. \ref{Sk}. The length scale was normalized by the smallest particle size $r_{\rm min}$. The structure factors decreased in a power-law in the range $k<10^{0.3}$, followed by a plateau region. The crossover of the two regions corresponds to the length around $2\pi/10^{0.3}\simeq 3$, which is close to the diameter of the smallest particles, $2r_{\rm min}$. The exponent in the decreasing region was found to be $-a+1$ and to deviate from it as $a$ approached 3 (inset of Fig. \ref{Sk}). The exponent of the structure factor $-a+1$ is also suggested theoretically in one dimension and numerically in two dimensions to appear in the densest packing of circles following a power size distribution\cite{cherny2023dense}.
The slope of the structure factor $-a+1$ is -1 times the fractal dimension of the densest packing, suggesting fractality.
However, the deviation from $-a+1$ when $a$ is close to three may sensitively reflect the loss of fractality.
Our previous research has shown that scale invariance is violated when $a$ approaches three because the effect of the lower bound of the particle size distribution $r_{\rm min}$ becomes non-negligible. 

\subsection{Packing fraction}

\begin{figure}
    \centering
    \includegraphics[width=56mm, bb = 0 0 431 343]{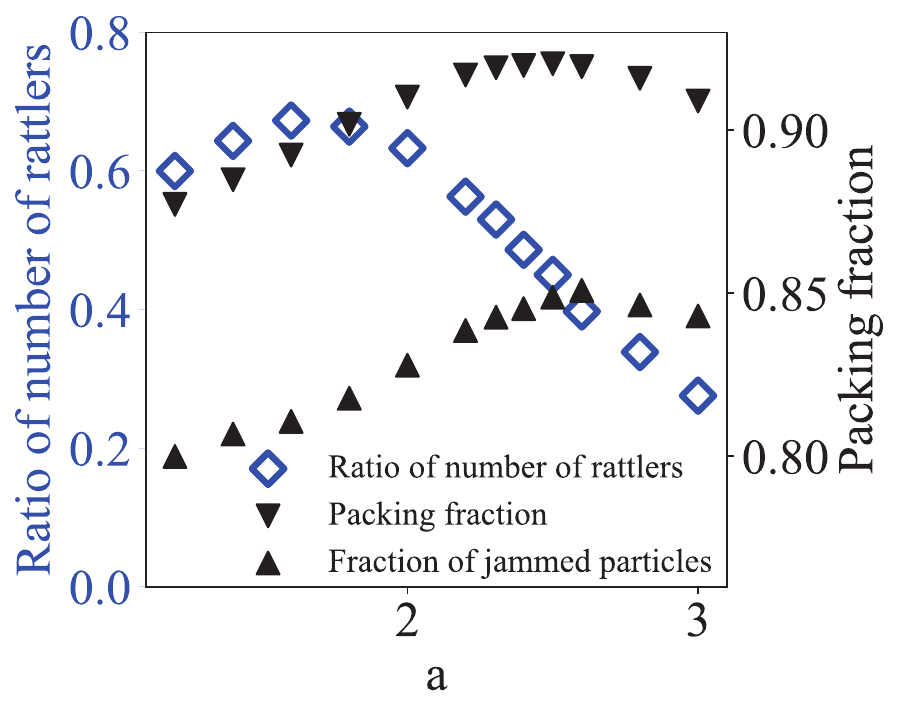}
    \caption{Downward pointing triangles: Dependence of packing fraction, $\phi_{\rm c}$ on $a$. Upward pointing triangles: Dependence of packing fraction except rattlers, $\phi_{\rm J}$ on $a$. White squares: Fraction of the number of rattlers in the total particles.}
    \label{packing_fraction}
\end{figure}

In our previous report \cite{shimamoto2023common}, we found that the jamming transition point, $\phi_{\rm J}$ has a maximum value of $\simeq 0.92$ for $2<a<3$. Here, we investigate the jammed structure from the analysis of the number ratio of rattlers, packing area fraction of all particles, $\phi_{\rm c}$, and packing fraction of particles excluding rattlers, $\phi_{\rm J}$ (Fig. \ref{packing_fraction}). 
For $a<2$, the number of rattlers has a maximum value of $>0.6$, but the area fraction occupied by rattles, $\phi_{\rm c}-\phi_{\rm J}$, is at most from $7\%$ to $8\%$, and is also strongly independent of $a$.

In the $a$ range of interest in the present study ($1.2\leq a \leq 3$), $\phi_{\rm J}$ was larger than the value of the bidisperse system, $\simeq 0.84$.
Both $\phi_{\rm J}$ and the area fraction of jammed particles were particularly high in the range of $2<a<3$. The values of $\phi_{\rm J}$ are compared to the packing fraction of the densest configuration, $\phi_{\rm dens}$, which is called hexagonal packing for monodisperse particles and is known to be $\pi/\sqrt{12}\simeq 0.9069$. When $R$ is finite value, $\phi_{\rm dens}$ is obtained exactly only when the exponent $a_{\rm AP}$ is equal to the Apollonian packing:
\begin{eqnarray}
    \phi_{\rm dens}(R)=\frac{\int^R_1 dr r^{d}r^{a_{\rm AP}}}{\int^R_0 dr r^{d}r^{a_{\rm AP}}}.
\end{eqnarray}
In two-dimension, $a_{\rm AP}$ is known to be $2.305684\dots$ \cite{manna1991precise}. Under the present simulation condition with $R$ set to a finite value of 50, the highest packing fraction $\phi_{\rm dens}(50)$ is $\simeq0.933873$. However, in the numerically obtained jammed system, $\phi_{\rm c}$ was found to be smaller than this value (results). Although random packing is known to coincide with the densest packing in monodisperse systems\cite{luding2001liquid}, it was confirmed that the densest arrangement is not reached by random packing when $a=a_{\rm AP}$ \textcolor{black}{because of jamming}.
Here, $\phi_{\rm dens}$ is not unity because $R$ is set to a finite value. However, in the limit of $R\rightarrow\infty$ in the range $a_{\rm dens}<a<d+1$, $\phi_{\rm dens}$ is asymptotic to unity. Whether $\phi_{c}$ asymptotes to unity in the limit of $R\rightarrow\infty$ is a nontrivial interesting question. Because the gap is connected for higher dimension $d>2$, unlike $d=2$, a sufficiently small particle can move all over void, and the possibility that $\phi_{c}$ is unity cannot be ruled out, which corresponds to the absence of jamming transition.

Although the area fraction of the rattler is almost independent of the exponent $a$, a very large number of particles ($>60$ \%) became rattlers only in the region $a<2$. This indicates that the rattlers consist almost exclusively of small particles, i.e., mainly only large particles are arrested. This result is consistent with our previous result that the characteristic length scales of the jammed particles are $r_{\rm max}$ for $a\leq 2$ \cite{shimamoto2023common}. Note that Rattler's area fraction converges to a nontrivial value in the $R$-infinity limit, but the ratio of the number of rattlers is expected to asymptotically approach unity.

\subsection{Laguerre Voronoi tessellation}
\begin{figure}
    \centering
    \includegraphics[width=86mm, bb = 0 0 230 221]{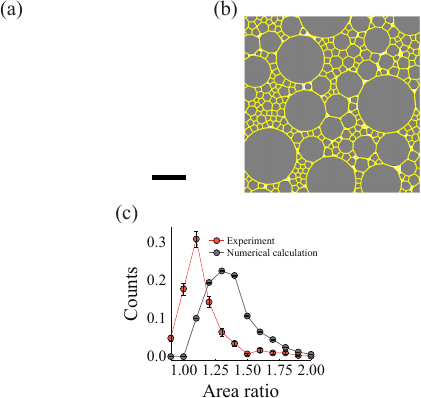}
    \caption{For the experiments (a) and the numerical calculations (b), the Laguerre Voronoi tessellation of each particle is overlaid on the packed particles as yellow polygons. Histograms of the ratio of the area of Laguerre Voronoi cell to the area of the corresponding particles, $f$ are shown in (c). The scale bar in (a) is 500 $\mu$m.}
    \label{voronoi_matome}
\end{figure}

Both experimentally and numerically obtained packings were analyzed by Laguerre Voronoi tessellation (for details on Laguerre Voronoi tessellation, see Appendix \ref{Laguerre}). Whole space was divided into cells, and the ratio of the area of a particle $A_{\rm p}$ to the area of the corresponding cell $A_{\rm c}$ was calculated as the ratio $f=A_{\rm c}/A_{\rm p}$ (Fig. \ref{voronoi_matome}). This area ratio $f$ is the area around each particle plus the size of the neighboring void, normalized by the area of the particle, and is the inverse of the local packing fraction.
The experimental results show a peak at $f=1.1$, while the numerical results show a peak at $f=1.3$.
At the jamming transition point $\phi_{\rm J}$, $f\geq 1$ must hold for all particles, but some cells with $f<1$ in the experimental results are owing to image processing errors.

\subsection{Diffence between experiments and simulations}
Both the experimentally and numerically constructed packing were in agreement in the contact network \cite{shimamoto2023common} and structure factors (Fig. \ref{Sk_exp}, \ref{Sk}). A difference was found in the probability distribution of the local packing fraction (Fig. \ref{voronoi_matome}).
The difference may be owing to the difference in the way the initial configuration was prepared and the optimization by annealing was applied only in the experiment.
The jamming particle systems that we used in both the experiments and the simulations had a common procedure in their preparations. They were generated by gradually increasing the packing fraction from a random initial configuration with a packing fraction slightly below $\phi_{\rm J}$. However, there is a clear difference between the two methods. In the numerical calculations, the particles were quenched from infinite temperature to zero temperature to reach a local minimum of energy. Whereas, in the experiments, the generated droplets have a more complex history than the particles used in the simulations, such as undergoing shear deformation during handling with a pipette or being placed between glass plates.

The influence of history on the structure of such jammed particles is already well known \cite{kumar2016memory,matsuyama2021geometrical}. Polydisperse systems are expected to be particularly strongly dependent on history, as is well known from Brazil's nut effect and rotating drum experiments \cite{oyama1939motion,zuriguel2005role,sanders2004brazil}. Therefore, a detailed analysis of the influence of history on the jammed structure of polydisperse particles is an important future challenge.

Another reason that can cause differences between experiments and simulations is the difference in the dimensions. In this study, we considered quasi-two-dimensional systems in the experiments and two-dimensional systems in the numerical calculations. The two-dimensional system differs from the three-dimensional and larger systems in that the contact points between particles separate the voids, i.e., the voids are not connected.
In three dimensions or more, particles that are sufficiently small can move around the void by passing through the space between the larger particles, which may have significantly different properties from those of two-dimensional systems.

\section{Conclusion}

First, we have explained the formation principle of oil droplets following the power size distribution by assuming that the formation process takes place under injection and multiplicative fracture.
This principle is expected to apply to a wide variety of materials. For quasi-two-dimensional droplets, the lower end of the particle size distribution is limited by the cell thickness and the upper end by the camera field of view, preventing extensive verification of the power law. This principle and the robustness of the distribution could be verified by solid materials whose size distributions are easier to measure.

Second, the packing structures at the jamming transition point, $\phi_{\rm J}$ for particles following the power size distribution $N(r)\propto r^{-a}$ were investigated experimentally and numerically. The number ratio of rattlers strongly depends on $a$, but the area fraction of the rattlers is almost independent of $a$. \textcolor{black}{Namely, in the region where $a$ is small, particles with a small area compose the majority of the rattlers, and large particles are mechanically constrained.}
Therefore, the representative length scale of the size of the jammed particles is larger in regions where $a$ is small.
From the analysis of structure factors, we find isotropicity of extremely polydisperse particles and power-like decay similar to that reported for more densely packed, above-jamming point, particles \cite{cherny2023dense}.

Particles with extremely large size distributions, which are the focus of this study, are the counterparts of monodisperse systems, and we have clarified their static structure from several points of view.
The qualitative discrepancy between experimental and numerical calculations may be owing to the history of the jamming transition point.
In the future, the effect of history on the structure of packing should be investigated.

\appendix
\section{Lognormal distribution caused by multiplicative fracturing \label{ap_lognormal}}

The random multiplication process leads to a lognormal distribution. We introduce the process of the derivation briefly based on Kolmogorov's result \cite{kolmogorov1941logarithmically}. Consider multiplying the variable $x_0$ by the multiplication noise $b_1, b_2,\dots b_m$ subsequently. Let $x_m$ be the value obtained by multiplying $x_0$ by $b_1$ to $b_m$ as follows:
\begin{subequations}
\begin{align}
    &x_m=b_mx_{m-1},\label{randomjozan}\\
    &x_m=x_0\prod_{i=1}^m b_i,
\end{align}
\end{subequations}
where the population of $b_i$ is assumed to be independent and identically distributed such that $\log b_i$ has mean $\braket{\log b_i}=\mu$ and variance $\sigma^2$. Considering the logarithm of both sides, it can be written as follows:
\begin{equation}
    \log{x_m} = \log{x_0}+\sum_{i=1}^m \log{b_i},
\end{equation}
where $m\gg1$. By the central limit theorem, the probability distribution of $P\log x_m$, $P(\log x_m)$ of $\log x_m$ converges to a normal distribution with mean $m\mu$ and variance $m\sigma^2$:
\begin{equation}
    P(\log x_m)=\frac{1}{\sqrt{2\pi m\sigma^2}}\exp\left(-\frac{\left(\log x_m-m\mu\right)^2}{2m\sigma^2}\right).
\end{equation}
By converting the variable, we obtain
\begin{eqnarray}
    P(x_m)&=&P(\log x_m)\frac{d\log x}{dx},\\
    &=&\frac{1}{\sqrt{2\pi m\sigma^2}x_m}\exp\left(-\frac{\left(\log x_m-m\mu\right)^2}{2m\sigma^2}\right). \label{lognormal}
\end{eqnarray}
Thus, the lognormal distribution was derived owing to the random multiplicative process.

\section{Laguerre Voronoi tessellation\label{Laguerre}}
The Laguerre Voronoi tessellation is a generalized Voronoi tessellation, also known as a power diagram. In a Voronoi tessellation, the cell corresponding to a site $p$ is defined as the set of points $x$ that satisfy $|x-p|^2<|x-q|^2$ for points $x$ in space and all sites $q$. In Laguerre Voronoi tessellation of a $d$-dimensional space $R^d$ and a set of sites $Q$, each site $p\in Q$ is assigned a weight $w(p)$, and the cell corresponding to the site $p$ is defined as the points $x$ that satisfy $|x-p|^2-w(p)\leq|x-q|^2-w(q)$ for all $q\in P$. In particular, setting $w(p)$ to $0$ for all $p\in P$ yields a Voronoi tessellation; the Laguerre Voronoi tessellation in dimension $d$ is also obtained as a cross-section by the $d$-dimentional plane $\Pi$ of the Voronoi tessellation in dimension $d+1$. Here, the sites of the Voronoi tessellation are expressed as the points perpendicularly $\sqrt{M-w(p)}$ away from the corresponding sites $p$ on $\Pi$, where $M$ is an arbitrary constant greater than $\max_{p\in{P}}w(p)$.

For non-overlapping polydisperse spheres, the local density can be calculated by considering Laguerre Voronoi tessellation. The sites are set to be the centers of the spheres, and the weights $w(p)$ are set to $r(p)^2$, and the tessellation is performed. The tessellation is expressed such that the cells encompass the  corresponding spheres and do not overlap with the other spheres. Here, we denote by $r(p)$ the radius of the sphere centered at the point $p$. A quantity reflecting the local density $f=A_{\rm c}/A_{\rm p}$ is obtained as the ratio of the volume of the corresponding cell $A_{\rm c}$ to the volume of the sphere $A_{\rm p}$.

\section{Controlling the size range of the droplets \label{size_dist}}
\begin{figure}[bhp]
    \centering
    \includegraphics[width=86mm, bb = 0 0 565 470]{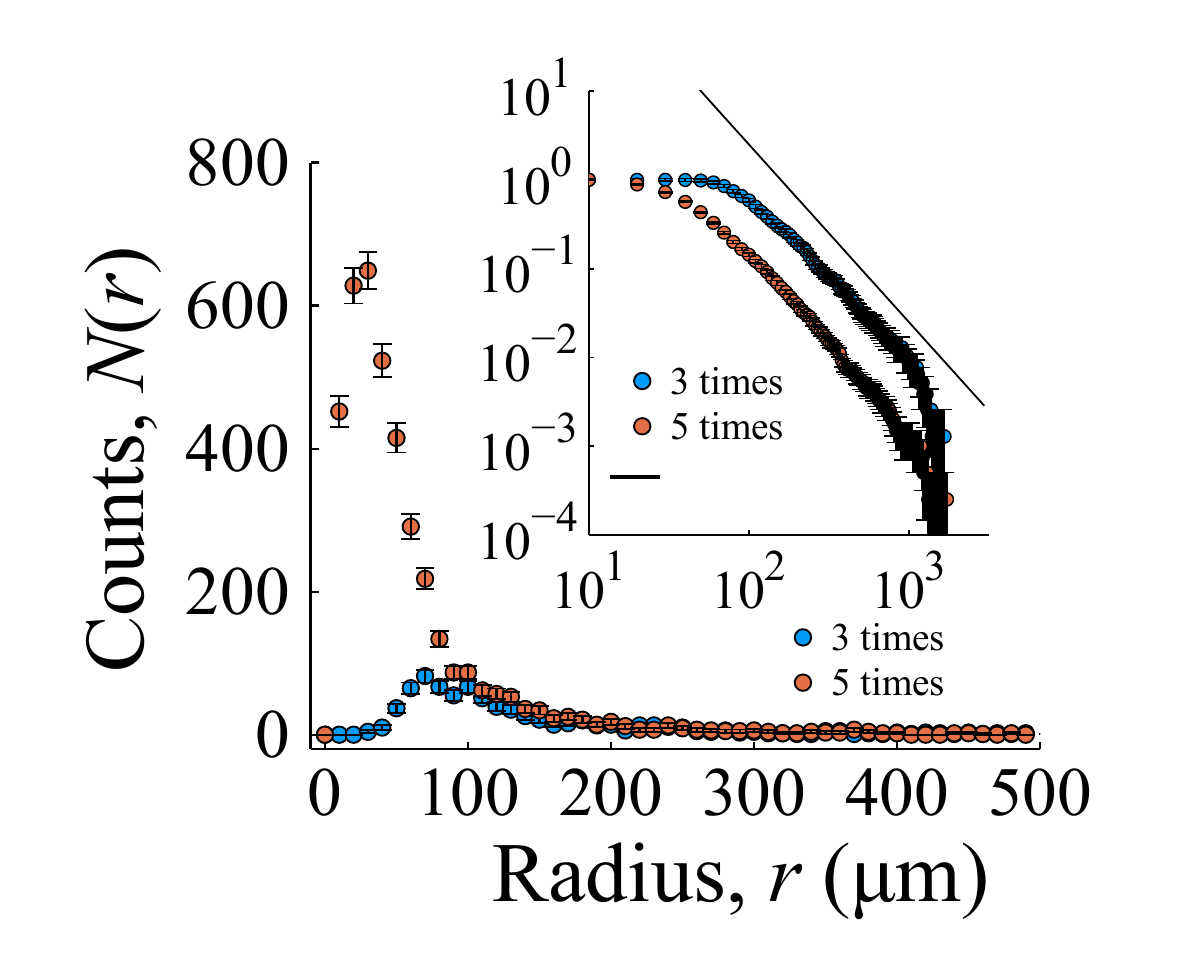}
    \caption{The particle size distribution of the quasi-two-dimensional droplets prepared by repeating injection and impact fracture. The number of repetitions was set to three and five. Note that the droplets with a radius of less than 45 nm are no longer quasi-two-dimensional because their diameter is less than the cell's thickness. The results were almost equivalent to Fig. \ref{disp_exp}(f) when the repetitions were 3 times.}
    \label{psd}
\end{figure}

Droplets were prepared by repeated fracture and injection in this study. We found that this method can control the range of the particle size distribution by changing the number of repetitions, but cannot control the index of the particle size distribution.
The results for three and five repetitions of destruction and injection are shown in Fig.\ref{psd}. The experimental procedure is almost the same as in the main text, but a tube was dropped from a height of 20 cm instead of tapping. This method is useful for comparing size distributions because it reduces fluctuations in the strength of each impact. Although the slope of the distribution did not change when the number of repetitions was 3 and 5, the range of particle size was found to increase as the number of repetitions increased.

\bibliography{poly_abrv2}

\end{document}